\documentclass[pra,twocolumn]{revtex4}

\usepackage{optidef}

\usepackage{graphicx,bm,natbib,upgreek,amsmath,mathrsfs,accents}
\usepackage{amsbsy}
\usepackage{amsfonts}

\usepackage{amsthm}

\usepackage[dvipsnames]{xcolor}

\usepackage{color}
\usepackage{enumitem}

\begin{document}

\title{Resource-efficient energy test and parameter estimation \\ 
in continuous-variable
quantum key distribution}

\author{Cosmo Lupo}
\affiliation{Department of Physics and Astronomy, The University of Sheffield, Sheffield, S3 7RH, United Kingdom.}

\begin{abstract}
Symmetry plays a fundamental role in the security analysis of quantum key distribution (QKD). 
Here we review how symmetry is exploited in continuous-variable (CV) QKD to prove the optimality of Gaussian attacks in the finite-size regime.
We then apply these results to improve the {feasibility} and the key rate of these protocols.
First we show how to improve the {feasibility} of the energy test, which is one important routine aimed at establishing an upper bound on the effective dimensions of the otherwise infinite-dimensional Hilbert space of CV systems.
Second, we show how the routine of parameter estimation can be made resource efficient in measurement-device independent (MDI) QKD. 
These results show that all the raw data can be used both for key extraction and for the routines of energy test and parameter estimation.
Furthermore, the improved energy test does not require active symmetrization of the measured data, which is computationally demanding.
\end{abstract}

\maketitle

\section{Introduction}

Quantum key distribution (QKD) is the art of exploiting quantum optics to generate a secret key between two distant users linked by an insecure quantum communication channel \cite{Rev1}. 
Once the secret key is established, it can be used as a one-time pad to achieve ever-lasting secure communication.
This is in contrast with public-key cryptography, including post-quantum cryptography, whose security relies on computational complexity and is thus undermined by increasing computational power, as well as quantum computing \cite{MMosca}.

QKD protocols can be classified according to the nature of the degrees of freedom that are used to encode information in the quantum electromagnetic field. 
Discrete-variable (DV) protocols encode information in a discrete way (often binary) using, for example, polarisation or time delay.
Continuous-variable (CV) protocols encode information in continuous degrees of freedom, for example quadrature and phase.

CV QKD has the strategic advantages of requiring, for decoding, off-the-shelf components as homodyne or heterodyne detection, which are mature technologies developed in standard tele-communication, whereas DV QKD needs high-efficiency and low-noise single-photon detectors \cite{Lo}. 
{For recent experimental demonstrations of CV QKD, see for example, Refs.\ \cite{Exp1,Exp2,Exp3}.}
However, CV QKD is still not completely characterized from a theoretical point of view. 
Indeed, technical issues make the mathematical analysis of CV QKD more challenging than DV QKD. This is essentially due to the fact that the quantum states prepared and shared in CV QKD live into a Hilbert space of infinite dimensions.

The security of CV QKD has been first established for \textit{Gaussian attacks}, i.e., under the assumption that the quantum communication channel maps Gaussian states into Gaussian states \cite{Gaussian}. 
Second, it has been extended to \textit{collective attacks}, i.e., assuming that each signal passing through the quantum channel is subject to an instance of the same noisy transformation \cite{Garcia,Navascues,Lev2015}.
The most general attacks are the so-called \textit{coherent attacks}, where the noisy communication channel can act in an unconstrained and unstructured way on the input signals. 
Finally, the highest standard of security is that of composable security, where the security of QKD is quantified and not only assessed in a qualitative way \cite{compo}.
The composable quantification of security is especially important in the finite-size scenario, where a finite number of quantum signals are exchanged, as it is the case in any experimental realisation of QKD.

For CV QKD, composable security against coherent attacks has been established for two kinds of protocols: 
the first is based on squeezed states and homodyne detection, whose security analysis exploits the phase-quadrature uncertainty relations \cite{Furrer,FurrerRR}; 
the second is based on coherent states and heterodyne detection and exploits the \textit{post-selection method} \cite{posts} to prove the optimality of Gaussian attacks in the finite-size regime \cite{Lev2017}.
In this paper we focus on the latter approach, which applies to \textit{no-switching} protocols introduced by Weedbrook et al.\ \cite{Weedbrook}.
In both cases the protocols include an \textit{energy test} to project the quantum state into a finite dimensional Hilbert space. 
The test requires to measure part of the system. Depending on the output of the measurement, one can assume that the rest of the system belongs, with high probability, to a Hilbert space of finite dimensions.
This implies that part of the raw keys, i.e., the data used for the energy test, cannot be used for key extraction.
Here we show how this limitation can be removed by applying an idea first introduced by Leverrier \cite{Lev2015}. 
The impact of this result on practical CV QKD is not on improving the key rate, but in making the protocol more experimentally feasible. In fact, our approach does not require Alice (the sender) and Bob (the receiver) to symmetrize their measured data \cite{Lev2017}. Such symmetrization is obtained by applying a random unitary rotation on their raw keys. Due to the large size of the data set (typically $10^8$-$10^{10}$ samples may be collected), the symmetrization is computationally demanding.
We remark that the analysis of Ref.\ \cite{Ghorai2019}, which also focused on the active symmetrization routine, did not consider its application to the energy test. 
\color{black}

Parameter estimation is a sub-routine of QKD whose goal is to obtain information about the communication channel.
The standard way of performing parameter estimation requires Alice and Bob to publicly announce part of their local raw keys. This implies that the data used for parameter estimation is compromised and cannot be used for key extraction, though there exist ways to bypass this limitation \cite{Lev2015}.
Here we consider parameter estimation in measurement-device-independent (MDI) QKD, and show that in this case all the raw data can be used for both parameter estimation and secret key extraction.

MDI QKD was introduced to avoid side-channel attacks on the measurement apparatus \cite{MDI1,MDI2}. {Experimental implementations are demanding but feasible, see for example Refs.\ \cite{MDI-exp1,MDI-exp2}}
The theory of Leverrier \cite{Lev2017} allows us to prove the composable security in the finite-size regime for a class of protocols based on preparation of coherent state and heterodyne detection, including MDI protocols \cite{Ghorai2019}. 
Recently, Refs.\ \cite{PRA2018,PRL2018} introduced modified CV MDI protocols that allow Alice and Bob to use all the measured data for both parameter estimation and key extraction. 
The scope of this result was questioned by Ghorai, Diamanti, and Leverrier \cite{Ghorai2019}, who suggested that it might \textit{not} be compatible with composable security.
Here we address this criticism and show that the protocol is indeed secure in a composable way, as follows from the application of the general method put forward in Ref.\ \cite{Lev2017}.
We thus establish that, in CV MDI QKD, all the raw data can be used for both key extraction and parameter estimation without compromising the composable security of the protocol. 
This is expected to have an impact in increasing the key rate in the finite-size regime.

While our result on parameter estimation is confined to MDI protocols (although in principle this can be extended to one-way protocols \cite{PRL2018}), our findings about the energy test may be as well applied to a wider range of CV QKD protocols.  
This is suggested by Ref.\ \cite{Ghorai2019}, where it was shown that the proof techniques of \cite{Lev2017} apply to two-way protocols with Gaussian displacements \cite{QZ2018-2way,Pirs2008}, as well floodlight QKD \cite{QZ2016,QZ2017,QZ2018}.
\color{black}


The paper proceed as follows. 
First we review the toolbox of CV quantum optics in Section \ref{sec:toolbox}, then we briefly review the use of symmetry in CV QKD in Section \ref{sec:Lev2017}. 
In Section \ref{sec:test} we introduce the resource-efficient energy test, where all the raw keys can be used for both the test and secret key extraction.
In Section \ref{sec:MDI_0} we introduce the CV MDI protocol in the entanglement-based representation, and in Section \ref{sec:MDI-PM} we consider its prepare \& measure representation.
Finally, in Section \ref{sec:debunk} we show that all the raw data can be used for both parameter estimation and key extraction in CV MDI QKD.
Conclusions are presented in Section \ref{sec:end}.

\section{The toolbox of continuous-variable quantum cryptography}\label{sec:toolbox}

In CV QKD information is encoded in the phase and quadrature of the quantum electromagnetic field \cite{Ferraro}.
The building block of CV QKD is the bosonic mode, also called \textit{qumode}, which is formally represented as a quantum harmonic oscillator with annihilation and creation operators, $\hat a$ and $\hat a^\dag$, satisfying the canonical commutation relations $[ \hat a , \hat a^\dag] = 1$.
%
The phase and quadrature operators are defined as $\hat q = (\hat a + \hat a^\dag)/\sqrt{2}$ and $\hat p = (\hat a - \hat a^\dag)/\sqrt{2}i$ respectively. 
The Hamiltonian of the harmonic oscillator is $\hat H = (\hat q^2 + \hat p^2)/2 = \hat a^\dag \hat a + 1/2$, where $\hat N = \hat a^\dag \hat a$ is the number operator. The latter has eigenvectors $\{ |n\rangle \}_{n=0, \dots, \infty}$ such that $\hat N |n\rangle = n |n\rangle$, which form a complete set, $\sum_{n=0}^\infty |n\rangle \langle n| = 1$. The number states can be obtained by repeated applications of the creation operator on the vacuum state as $|n\rangle = \frac{1}{\sqrt{n!}} \left( a^\dag \right)^n |0\rangle$, with $\hat N |0\rangle = 0$.

Here we focus on QKD protocols where classical information is encoded in a qumode in a continuous way using coherent states. A coherent state $|\alpha\rangle$ is an eigenvector of the annihilation operator, $\hat a |\alpha\rangle = \alpha |\alpha\rangle$, where $\alpha \in \mathbb{C}$ is a complex amplitude. Coherent states are expanded in the number basis as $|\alpha\rangle = e^{-|\alpha|^2/2}\sum_{n=0}^\infty \frac{\alpha^n}{\sqrt{n!}} |n\rangle$. They form an over-complete set that satisfies $\frac{1}{\pi} \int d^2\alpha |\alpha\rangle \langle \alpha| = 1$, where $d^2 \alpha = d \mathrm{Re}(\alpha) d \mathrm{Im}(\alpha)$.
In CV QKD, a pair of classical random variables $q^\mathrm{pre}, p^\mathrm{pre} \in \mathbb{R}$ are encoded in the coherent state $|\alpha\rangle$ with amplitude $\alpha = (q^\mathrm{pre} + i p^\mathrm{pre})/\sqrt{2}$.

Heterodyne detection is formally represented as the continuous family of POVM elements $\Lambda_\beta = \frac{1}{\pi} |\beta \rangle \langle \beta|$, with $\beta \in \mathbb{C}$. 
We have $\mathrm{Tr} \left( \Lambda_\beta |\alpha\rangle \langle \alpha| \right) = \frac{1}{\pi} |\langle \alpha | \beta \rangle |^2 = \frac{1}{\pi} e^{- | \alpha - \beta|^2}$.
Given a quantum state $\rho$, the output of heterodyne detection $\beta = (q^\mathrm{het} + i p^\mathrm{het})/\sqrt{2}$ defines the random variables  $q^\mathrm{het}, p^\mathrm{het} \in \mathrm{R}$.
An outcome of this measurement in the range $\beta \pm d\mathrm{Re}\beta/2 \pm i d\mathrm{Im}\beta/2$ has probability $\frac{1}{\pi} \, d^2\beta \mathrm{Tr} \left( \Lambda_\beta \rho \right) = \frac{1}{\pi} \,  d^2\beta  \langle \beta | \rho | \beta \rangle$.

The displacement operator is a unitary operator defined as $D(\gamma) = e^{ \gamma \hat a^\dag - \gamma^* \hat a }$. It allows us to shift the annihilation and creation operators by a c-number, i.e., $D(\gamma)^\dag \hat a D(\gamma) = \hat a + \gamma$, and $D(\gamma)^\dag \hat a^\dag D(\gamma) = \hat a^\dag + \gamma^*$.
It follows that the displacement operator also shifts the quadrature operators,
$D(\gamma)^\dag \hat q D(\gamma) = \hat q + \mathrm{Re}(\gamma)/\sqrt{2}$ and
$D(\gamma)^\dag \hat p D(\gamma) = \hat p + \mathrm{Im}(\gamma)/\sqrt{2}$.
The displacement operator maps coherent states into coherent states, $D(\gamma) |\alpha\rangle = |\alpha + \gamma\rangle$, and commutes with heterodyne detection, i.e., 
$\mathrm{Tr}\left( \Lambda_\beta D(\gamma) \rho  D(\gamma)^\dag \right)
= \mathrm{Tr}\left( D(\gamma)^\dag \Lambda_\beta D(\gamma) \rho \right) 
= \mathrm{Tr}\left( \Lambda_{\beta-\gamma} \rho  \right)$.
This means that displacing and then measuring is equivalent to measuring and then displacing the measurement output.

The two-mode squeezed vacuum (TMSV) is a quantum state of two qumodes. 
Its expansion in the number basis is 
\begin{align}
|\Psi \rangle_{AA'} = \sqrt{ \frac{1}{N+1} } \sum_{n=0}^\infty \left( \frac{N}{N+1} \right)^{n/2} |n\rangle_A |n\rangle_{A'} \, , 
\end{align}
where the parameter $N$ quantifies the mean photon number per qumode. 
We have 
\begin{align}
{}_{A}\langle \beta | \Psi \rangle_{AA'} 
& = \sqrt{ \frac{1}{N+1} }  e^{-\frac{1}{N+1} |\beta|^2/2} 
\left| \sqrt{ \frac{N}{N+1}} \beta^* \right\rangle_{A'}
\, .
\end{align}
This shows that if we measure one qumode of the TMSV $\Psi$ by heterodyne detection and obtain $\beta$, then the other qumode is prepared in the coherent state $|\alpha\rangle$ with amplitude $\alpha = \sqrt{ \frac{N}{N+1} } \, \beta^*$.
An outcome of this measurement in the range 
$\beta \pm d\mathrm{Re}\beta/2 \pm id\mathrm{Im}\beta/2$ 
has probability 
$d^2\beta \frac{1}{\pi} \frac{1}{N+1} e^{-\frac{1}{N+1} |\beta|^2}
=
d^2\alpha \frac{1}{\pi} \frac{1}{N} e^{-\frac{1}{N} |\alpha|^2}$.

Consider a system of $\ell$ qumodes and the $2\ell$ classical phase and quadrature variables $\mathbf{q} = q_1, \dots q_\ell$, $\mathbf{p} = p_1, \dots p_\ell$.
For any density operator $\rho$, the Wigner function is defined as
\begin{align}
    W(\mathbf{q},\mathbf{p}) = \frac{1}{\pi^\ell} \int_{-\infty}^\infty d^n \mathbf{y} \langle \mathbf{q} + \mathbf{y} | \rho | \mathbf{q} - \mathbf{y} \rangle e^{-2 i \mathbf{p} \cdot \mathbf{y}} \, ,
\end{align}
where $\mathbf{y} = y_1, \dots y_\ell$, $d^n\mathbf{y} = dy_1 \dots dy_\ell$, and $|\mathbf{q} \pm \mathbf{y}\rangle$ are eigenvectors of the quadrature operators, i.e., 
$\hat q_j |\mathbf{q} \pm \mathbf{y}\rangle = (q_j \pm y_j ) |\mathbf{q} \pm \mathbf{y}\rangle$.
By definition, a Gaussian state has a Wigner function that is a multivariate normal distribution.
Gaussian states are therefore uniquely determined by the first moments and the (symmetrically ordered) covariance matrix (CM) of the quadrature operators.

\section{Symmetry in CV QKD}\label{sec:Lev2017}

In general, the experimental realisation of the protocol follows the \textit{Prepare \& Measure} (PM) representation, where the legitimate users prepare quantum states and send them through an insecure quantum channel controlled by the eavesdropper (Eve). 
For example, in one-way protocols Alice sends quantum signals to Bob, who measures them, and in MDI protocols both Alice and Bob send signals to a central relay.
However, the security of the protocol is proven in the equivalent \textit{Entanglement-Based} (EB) representation, where a bipartite quantum state $\rho_{AB}^n$ is distributed to Alice and Bob, and Eve holds a purification.
If the protocol is secure in the EB representation, so it is in the PM one.
For this reason, in this section we review security proof of Leverrier \cite{Lev2017}, which is defined in the EB representation.

A QKD protocol typically acts on $n$ instances of a given physical system. 
CV QKD, in the EB representation, is defined on a set of $2n$ qumodes.
For $j=1,\dots, n$, these are represented by the creation and annihilation operators $a_j^\dag$, $a_j$, and $b_j^\dag$, $b_j$, which are associated to Alice and Bob respectively.

An EB QKD protocol is formally associated to a completely positive (CP) map $\mathcal{E}$ that takes as input $\rho_{AB}^n$ and outputs a shared key. 
The latter is represented by a classical state of the form 
$\sum_{x,y=0}^{2^\ell-1} p(x,y) |x\rangle_A \langle x | \otimes |y\rangle_B \langle y|$, 
where $x,y=0,\dots,2^\ell-1$ are the possible keys obtained by Alice and Bob, respectively.
In reality, we expect the keys to be only approximately secret, as imperfections in the protocol may leak information to Eve.
To assess the security of a QKD protocol we compare $\mathcal{E}$ with the ideal map $\mathcal{E}_0$ that takes any input state and replace it with a perfectly secret key of $\ell$ bits. 
The latter is represented by a classical-quantum state $\sum_{x=0}^{2^\ell-1} 2^{-\ell} |x\rangle_A \langle \otimes |x\rangle_B \langle x| \otimes \rho_E$ where Eve has no information about the key.
We then define $\Delta := \mathcal{E} - \mathcal{E}_0$, and require that
\begin{align}
    \| \Delta \|_\diamond \leq \epsilon \, ,
\end{align}
for some $\epsilon \ll 1$.
We recall that the diamond norm, $\| \, \cdot \, \|_\diamond$, is defined as the worst-case trace norm over all possible input states, including their purification,
\begin{align}
    \| \Delta \|_\diamond 
    := \sup_\psi \| (\Delta \otimes I)  \psi_{ABE} \|_1 \, ,
\end{align}
where the maps $\Delta$ acts on the systems $AB$ associated to Alice and Bob, and $I$ is the identity map on the purifying system $E$. 
Finally, the supremum is over all tripartite states $\psi_{ABE}$, and $\| O \|_1 := \mathrm{Tr}|O|$ is the trace-norm.

The security analysis is thus reduced to the task of estimating the above diamond norm. 
The upper bound $\epsilon$ quantifies the (in)security of the protocol, the smaller $\epsilon$ the more the protocol is secure. Operationally, the \textit{security parameter} $\epsilon$ quantifies the probability to discriminate between the actual protocol $\mathcal{E}$ and the ideal one $\mathcal{E}_0$ \cite{PRA2018}. 
Estimating the diamond norm is a very challenging task as it requires the calculation of the supremum over all states in a high dimensional Hilbert space. In the case of CV QKD this space is infinite-dimensional. 
Fortunately, the task can be dramatically simplified by exploiting symmetry. 

CV QKD protocols that are based on heterodyne detection may be invariant under a symmetry group of \textit{passive linear optics} transformations that mix the qumodes but preserves the separation between Alice and Bob.
For any $n \times n$ unitary matrix $U$, consider the linear transformation on the bosonic operators, $a_j \to \sum_{k=1}^n U_{jk} a_k$ and $b_j^\dag \to \sum_{k=1}^n U_{jk} b_k^\dag$. 
This map defines a bosonic representation $R$ of the group $\mathrm{U}(n)$ of $n \times n$ unitary matrices \cite{Aniello}, 
\begin{align}
a_j & \to R_U a_j R_U^\dag = \sum_{k=1}^n U_{jk} a_k \, , \\
b_j^\dag & \to R_U b_j^\dag R_U^\dag = \sum_{k=1}^n U_{jk} b_k^\dag \, .
\end{align}
The protocol is covariant under this group of transformation if for any unitary $U$ there exists a CPT map $K_U$ such that 
\begin{align} \label{covariance}
\Delta \circ R_U = K_U \circ \Delta \, .
\end{align}
Under this condition it is easy to show that \cite{posts}
\begin{align}\label{diamond-2}
    \| \Delta \|_\diamond 
    & = \sup_{\bar\rho_{AB}^n \in \mathcal{F}_n} \| (\Delta \otimes I) \bar\rho_{ABE}^n \|_1 \, ,
\end{align}
where $\mathcal{F}_n$ is the subspace of states that are invariant under the symmetry group, i.e., $R_U \bar\rho_{AB}^n R_U^\dag = \bar\rho_{AB}^n$.
The subspace $\mathcal{F}_n$ is spanned by the $\mathrm{SU(1,1)}$-coherent states over $2n$ qumodes \cite{GdeFinetti}.
The $\mathrm{SU(1,1)}$-coherent states have a number of important properties, in particular (1) they are Gaussian state and (2) provide a decomposition of the unity.

Equation (\ref{diamond-2}) cannot yet be used to compute the diamond norm. To further simplify it, we need to prepend a suitable \textit{energy test} $\mathcal{T}$ to the QKD protocol $\mathcal{E}$. 
The energy test consists in measuring $2k < 2n$ qumodes by heterodyne detection. According to the measurement result, the remaining $2n-2k$ qumodes are projected, up to a small probability of error $\epsilon_\mathrm{test}$, into a finite-dimensional Hilbert space. 

Conditioned on passing the energy test, the state of the remaining $2n-2k$ qumodes can be effectively assumed to live in a finite dimensional Hilbert space.
This allows us to write 
\begin{align}\label{post-d}
    \| \Delta \circ \mathcal{T} \|_\diamond \leq 
    c_{n-k,d} \sup_{\sigma_{ABE}} \| (\Delta \otimes I) \sigma_{ABE}^{\otimes n-k} \|_1 + \epsilon_\mathrm{test} \, ,
\end{align}
where the supremum is over Gaussian states $\sigma_{ABE}$, and $c_{n-k,d} = K^4/50$, and $K \sim (n-k) ( d_A + d_B)$ gives a bound on the effective dimesions.
In conclusion, this shows that a protocol that is secure against Gaussian attacks is also secure against general coherent attacks, provided that the energy test is passed, and than one is willing to pay a multiplicative penalty in the security parameter.

The protocol $\mathcal{E}$ is composed of several sub-routines. 
After the energy test $\mathcal{T}$, one proceeds with the measurement $\mathcal{M}$ (heterodyne detection for this class of symmetric protocols),
then the parameter estimation routine $\mathcal{P}$, followed by error correction $\mathcal{C}$ and privacy amplification $\mathcal{A}$. We can therefore write $\mathcal{E}  \circ \mathcal{T} = \mathcal{A} \circ \mathcal{C} \circ \mathcal{P} \circ \mathcal{M} \circ \mathcal{T}$. 
Since the energy test and the measurement commutes, we can as well write $\mathcal{E}  \circ \mathcal{T} = \mathcal{A} \circ \mathcal{C} \circ \mathcal{P} \circ \mathcal{T} \circ \mathcal{M}$. 
Following an argument put forward by Ghorai et al.\ \cite{Ghorai2019}, we do not need that the overall protocol $\mathcal{E} \circ \mathcal{T}$ is covariant.
In fact, it is sufficient that the covariance property (\ref{covariance}) holds for the energy test $\mathcal{T}$ and the parameter estimation $\mathcal{P}$.

\section{Resource-efficient energy test}\label{sec:test}

In this section we improve on the energy test introduced in Ref.\ \cite{Lev2017} to make it more resource efficient {and experimentally feasible}.
In Ref.\ \cite{Lev2017}, the energy test applies to a bipartite state $\rho_{AB}^n$ of $2n$ qumodes, where $n$ qumodes are on Alice side, and $n$ on Bob side:
the result of heterodyne detection on $2k < 2n$ qumodes allows Alice and Bob to establish an upper bound on the dimensions of the Hilbert space containing the state of the remaining $2(n-k)$ qumodes.
The qumodes measured for the energy test cannot be used for key extraction, therefore, reducing their number from $2n$ to $2(n-k)$.

First we review the energy test of Ref.\ \cite{Lev2017}. Then we modify it to make it resource efficient, in such a way that all the qumodes are used for both the energy test and key extraction.
The energy test of Ref.\ \cite{Lev2017} is defined as follows:
\begin{itemize}
    \item Alice and Bob publicly agrees on a random unitary matrix of size $n$ and apply the local transformations 
    $a_j \to \sum_{h=1}^n U_{jh} a_{h}$, 
    $b_j^\dag \to \sum_{h=1}^n U_{jh} b_{h}^\dag$.

    \item They measure by heterodyne detection the first $2k < 2n$ qumodes.
    The output of Alice's measurement is $\alpha_1 \dots \alpha_n$, and the output of Bob's measurement is $\beta_1 \dots \beta_n$.
    
    \item They compute the quantities $E_A^k = \frac{1}{k} \sum_{j=1}^k |\alpha_j|^2$ and 
    $E_B^k = \frac{1}{k} \sum_{j=1}^k |\beta_j|^2$.

    \item If $E_A^k \leq d_A$ and $E_B^k \leq d_B$ for some $d_A$, $d_B$, then they conclude that the state of the remaining $2(n-k)$ qumodes lives in a Hilbert space of dimensions not larger than
    $K = (n-k) \left( d_A' + d_B' \right)$,
    where 
    $\star = A, B$,
    $d_\star' =  d_\star \, g(n-k,k,\epsilon/4)$, and
    \begin{align} \label{mbx3x3}
    g(n_1,n_2,\delta) = \frac{1 + 2 \sqrt{\frac{\ln{(2/\epsilon)}}{2n_1}} + \frac{\ln{(2/\epsilon)}}{n_1}} {1-2\sqrt{\frac{\ln{(2/\epsilon)}}{2n_2}}} \, .
    \end{align}
    This statement holds with probability at least equal to $1-\epsilon$.

\end{itemize}

We now present a modified test that allows Alice and Bob to use all the modes for both the test and for key extraction.
We do that by applying an idea first introduced in Ref.\ \cite{Lev2015} to solve a similar problem in the context of parameter estimation. The test is applied after all qumodes have been measured by heterodyne detection.
The resource-efficient energy test is defined as follows:
\begin{itemize}
    
    \item Alice and Bob measure all their local qumodes by heterodyne detection, obtaining the output variables $\alpha_1 \dots \alpha_n$ and $\beta_1 \dots \beta_n$.

    \item They compute the quantities 
    $E_A^n = \frac{1}{n} \sum_{j=1}^n |\alpha_j|^2$ and 
    $E_B^n = \frac{1}{n} \sum_{j=1}^n |\beta_j|^2$.

    \item If $E_A^n \leq d_A$ and $E_B^n \leq  d_B$ for some integers $d_A$, $d_B$, 
    then they conclude that the original state $\rho_{AB}^{n}$ lived in a Hilbert space of dimensions not larger than
    $K = n \left( d_A' + d_B' \right)$, with $d'_\star = (1+\delta) d_\star$, and
    \begin{align} \label{wi3jq}
    \delta = 1.5 \sqrt{\frac{\ln{(2/\epsilon)}}{n/2}} \, .
    \end{align}
    As shown below, this statement holds with probability at least $1 - 4\epsilon$.
    
\end{itemize}

We now show that the resource-efficient energy test is in fact equivalent to the original energy test of Ref.\ \cite{Lev2017}. \\
\textbf{Proof:} First of all, we note that the quantities $E_A^n$, $E_B^n$ are invariant under the symmetry group.
Therefore, we can equivalently write 
$E_A^n = \frac{1}{n} \sum_{j=1}^n |\alpha_j'|^2$
$E_B^n = \frac{1}{n} \sum_{j=1}^n |\beta_j'|^2$, where
$\alpha_j' = \sum_{h=1}^n U_{jh} \alpha_{h}$ and $\beta_j = \sum_{h=1}^n U_{jh}^* \beta_{h}$,
for any unitary matrix $U$.
We can then define a thought experiment where Alice and Bob first randomize they measurement outcomes by applying a random unitary $U$, then apply the energy test of Ref.\ \cite{Lev2017} on the first
$n_1 = n/2$ qumodes to obtain an upper bound on the dimensions of the remaining $n_2 = n/2$ qumodes (for simplicity we assume that $n$ is even).
In this case, they would compute the quantities $E_A^{n_1}$, $E_B^{n_1}$.
Similarly, we could consider a dual thought experiment where they measure the last $n_2$ modes to bound the dimension of the first block of $n_1$ modes. In this case they would compute the quantities $E_A^{n_2}$, $E_B^{n_2}$. 
The two thought experiments are linked by the invariant relations $E_A^n = E_A^{n_1} + E_A^{n_2}$ and $E_B^n = E_B^{n_1} +E_B^{n_2}$.
To draw a conclusion about the two thought experiments from the knowledge of $E_A^n$, $E_B^n$, we can apply the following tail bound \cite{Lev2017}.
For a random choice of the unitary $U$, it holds that, for any $\epsilon \geq 2 e^{-n_1/2}$,
\begin{align}
\mathrm{Pr} \left\{ 
E_\star^{n_j} \geq \left[ 1 + 1.5 \sqrt{\frac{\ln{(2/\epsilon)}}{n/2}} \right] 
E_\star^n \right\} \leq \epsilon \, ,
\end{align}
for $\star = A, B$ and $j=1,2$.
This means that if $E_\star^n \leq d_\star$,
then with probability at least $1-\epsilon$, we also have $E_\star^{n_j} \leq d_\star (1+\delta)$ for a random choice of the unitary matrix $U$. 
This implies that both thought experiments would be successful with probability at least $1-4\epsilon$. $\Box$

In conclusion, at the price of paying an extra additive penalty of $4\epsilon$, the resource-efficient energy test allows Alice and Bob to:
(1) use all their modes for both the energy test and for key extraction; and
(2) avoid the application of the transformation $a_j \to \sum_{h=1}^n U_{jh} a_{h}$, $b_j^\dag \to \sum_{h=1}^n U_{jh} b_{h}^\dag$.

\begin{figure}[t!]
\includegraphics[width=0.9\linewidth]{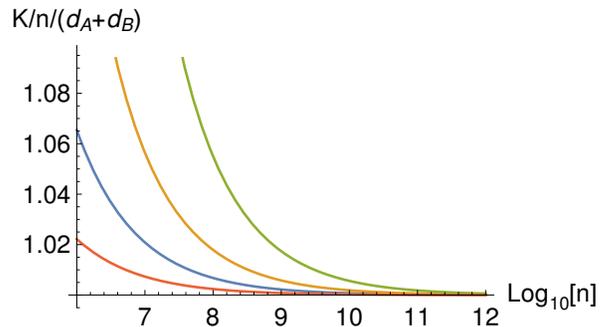}
 \caption{The normalized effective dimension $K/n/(d_A+d_B)$ of the truncated Hilbert space subject to passing the energy test, plotted vs the sample size $n$. 
 The three top lines are for the energy test of Ref.\ \cite{Lev2017}, with (from top to bottom
 $k = 10^{-3} n$ (green line), 
 $k = 10^{-2} n$ (orange line), and
 $k = 10^{-1} n$ (blue line).
 The bottom line (red) is for our efficient energy test that uses all data for both the test and for key extraction. 
 The lines are obtained by imposing that the probability of failing the test is smaller than $10^{-20}$.}
\label{fig:Ks}
\end{figure}

In summary, the goal of the energy test is to estimate an effective, finite dimension for the Hilbert space of Alice and Bob signals. 
In the limit of $n\to\infty$ (and $k \to \infty$ for the energy test of Ref.\ \cite{Lev2017}) the effective dimension, per symbol sent, is given as $K/n = d_A + d_B$.
For finite $n$ and $k$, the estimate for this dimension is larger by a multiplicative factor $K/n/(d_A+d_B)$. 
For the energy test of \cite{Lev2017}, the factor is $g(n-k,k,\epsilon/4)$, from Eq.\ (\ref{mbx3x3}). For our energy test it is $1+1.5 \sqrt{2\ln{(8/\epsilon)}/n}$, from Eq.\ (\ref{wi3jq}).
Figure \ref{fig:Ks} compares the ratio $K/n /(d_A + d_B)$ for both approaches.
%
%
The fact that they are both of order one shows that, as anticipated, our efficient energy test is not substantially changing the key rate in the finite-size regime (though a small improvement is observed). 
As discussed above, the important impact of our energy test is to make the protocol experimentally feasible.
\color{black}

\section{CV MDI QKD}\label{sec:MDI_0}

Measurement-device independent (MDI) QKD was introduced to avoid side-channel attacks on the measurement apparatus \cite{MDI1,MDI2}. 
In the PM representation, the legitimate users Alice and Bob prepare and send quantum states, whereas the task of measuring them is delegated to a third, untrusted party. Here we refer to the latter as the relay, and assume without loss of generality that it is fully controlled by Eve.
On the contrary, in the EB representation Alice and Bob also need to apply some local measurements. However, this does not expose MDI to side-channel attacks as the EB representation is only used as a theoretical tool to prove the security of the equivalent PM protocol. It is only the PM protocol that is implemented experimentally.

It is useful to formally split the description of the MDI protocol in two parts. 
The first part is \textit{state preparation}, whose goal is to establish a correlated quantum state $\rho_{AB}^n$ shared between the legitimate users. 
It is the preparation phase that characterizes CV MDI QKD with respect to other one-way and two-way CV QKD protocols.
Once the quantum state has been distributed, the users proceed with \textit{key extraction}. 
The map $\mathcal{E}$ discussed in Section \ref{sec:Lev2017} corresponds to the key extraction part of the protocol, which takes the state $\rho_{AB}^n$ as input.
This second part of the MDI protocol comprises local measurements, parameter estimation, error correction, and privacy amplification. Here we consider an MDI protocol where the measurement is heterodyne detection.

The phase of state preparation, in the EB protocol, is as follows:
\begin{enumerate}

    \item Alice prepares a two-mode squeezed vacuum (TMSV) state $\psi_{AA'}$, with mean photon number $N_A$, on her modes $A$ and $A'$. 
    Similarly, Bob prepares a TMSV state $\psi_{BB'}$, with mean photon number $N_B$, on his local modes $B$ and $B'$.
    They retain modes $A$, $B$ and send the modes $A'$, $B'$ to the relay.
    
    \item The relay publicly announces a complex number $z = (q_Z + i p_Z)/\sqrt{2}$.
    
    \item Alice and Bob apply the phase-space displacement operators $D(\gamma_A)$, $D(\gamma_B)$ to their local modes, with displacement amplitudes determined by $z$, 
    \footnote{A more general displacement, of the form $\gamma_A = u_A z + v_A z^*$, $\gamma_B = u_B z^* + v_B z^*$, with $u_A, v_A, u_B, v_B \in \mathbb{C}$ was considered in Refs.\ \cite{PRA2018,PRL2018}. 
    The displacement rule in Eq.\ (\ref{aknucsdisp-1}) is less general but is explicitly covariant under the transformation 
    $a_j \to \sum_{k=1}^n U_{jk} a_k$, 
    $b_j^\dag \to \sum_{k=1}^n U_{jk} b_k^\dag$, 
    $z_j \to \sum_{k=1}^n U_{jk} z_k$ 
    for any unitary matrix $U$.
    }
    \begin{align}\label{aknucsdisp-1}
        \gamma_A = a z \, , \, \, 
        \gamma_B = b z^* \, ,
    \end{align}
    where $a$ and $b$ are real-valued constants, and $z^*$ denotes the complex conjugate of $z$.
\end{enumerate}
These steps are repeated $n$ times. The state prepared in this way is denoted as $\rho_{AB}^n$ and is the input of the second phase of the protocol. 
The phase of key extraction develops as follows:
\begin{enumerate}
  \setcounter{enumi}{3}
  
    \item Alice and Bob measure their local modes by heterodyne detection. The output is a pair of $n$-fold complex vectors, $\alpha = \alpha_1 \dots \alpha_n$ and $\beta = \beta_1 \dots \beta_n$, respectively, which represent their local raw keys.

    \item They perform the energy test, for example the resource-efficient energy test described in Section \ref{sec:test}.
    
    \item They apply a random rotation on their raw keys, $\alpha_j \to \sum_{h=1}^n U_{jh} \alpha_h$, and $\beta_j \to \sum_{h=1}^n U_{jh}^* \beta_h$, where $U$ is a $n \times n$ unitary matrix.
    
    They then select $k < n$ elements (for example $\alpha_1, \dots, \alpha_k$ and $\beta_1, \dots, \beta_k$) and share their values on a public channel. These data allow Alice and Bob to estimate the number of secret bits that can be distilled from what is left of the raw keys.
    This is discussed in more detail below.
    
    \item Alice and Bob perform error correction and privacy amplification to distill their secret keys from the remaining raw data of size $n-k$.

\end{enumerate}

In order to apply the theory of Ref.\ \cite{Lev2017} it is important to verify that the energy test and parameter estimation routine are covariant under the symmetry group.
The resource-efficient energy test is clearly covariant by construction. The same holds for the parameter estimation routine, as we now discuss in detail.
In this description of the protocol, we require active symmetrization of the local modes, though there are ways to avoid it \cite{Lev2015}.

\subsection{Covariant parameter estimation}

Alice and Bob share the quantum state $\rho_{AB}^n$, they measure it by local heterodyne detection, and obtain the raw data $\alpha= \alpha_1, \dots, \alpha_n$, $\beta= \beta_1, \dots, \beta_n$. 
The fundamental question of QKD is to determine how many secret bits they can extract from their raw data. 
According to the theory of Ref.\ \cite{Lev2017}, to answer this question we can without loss of generality assume that the state $\rho_{AB}^n$ is Gaussian, symmetric, and is factorized, i.e., $\rho_{AB}^n = \rho_{AB}^{\otimes n}$ 
(provided that the QKD protocol is covariant and the state has passed the energy test).

In other words, the initial state $\rho_{AB}^n$ can be assumed to be the collection of $n$ identical copies of a two-mode Gaussian state $\rho_{AB}$.
We recall that a Gaussian state is uniquely determined by the first and second moments of the quadrature operators, denoted as $\hat q_A$, $\hat p_A$ for Alice mode, and as $\hat q_B$, $\hat p_B$ for Bob mode. 
In Appendix \ref{App:CM} we review how we can obtain the CM of the Wigner function from the one of the outputs of heterodyne detection.
Using relation (\ref{wiem4}), the CM of the Wigner function can be obtained from the CM of the heterodyne measurement outputs in Eq.\ (\ref{734noi}).
To simplify the notation we drop the label “$\text{het}$” and indicate the heterodyne output variables simply as $q_A$, $p_A$, $q_B$, $p_B$.

Since the state is symmetric, we can further assume that the first moments $\mathbb{E}[ q_A ]$, $\mathbb{E}[ p_A ]$, $\mathbb{E}[ q_B ]$, $\mathbb{E}[ p_B ]$ vanish \footnote{The symmetry of the state implies that
$\mathbb{E}[ q_A ] + i \mathbb{E}[ p_A ] = \sum_k U_{jk} \left( \mathbb{E}[ q_A ] + i \mathbb{E}[ p_A ] \right)$ for any unitary matrix $U$. This condition is satisfied if only if $\mathbb{E}[ q_A ] + i \mathbb{E}[ p_A ] = 0$. The same conclusion holds for $\mathbb{E}[ q_B ]$ and $\mathbb{E}[ p_B ]$.}, and that the CM has the following form \cite{Lev2017}
\begin{widetext}
\begin{align}
    V_\mathrm{EB}^\mathrm{het} = \frac{1}{2} \left(
    \begin{array}{cccc}
    \mathbb{E}[ q_A^2 ] + \mathbb{E}[ p_A^2 ] & 
    0 & 
    \mathbb{E}[ q_A q_B ] - \mathbb{E}[ p_A p_B ] & 
    \star \\
    0 & 
    \mathbb{E}[ q_A^2 ] + \mathbb{E}[ p_A^2 ] & 
    \star & 
    - \mathbb{E}[ q_A q_B ] + \mathbb{E}[ p_A p_B ] \\
    \mathbb{E}[ q_A q_B ] - \mathbb{E}[ p_A p_B ] & 
    \star & 
    \mathbb{E}[ q_B^2 ] + \mathbb{E}[ p_B^2 ] & 
    0 \\
    \star & 
    - \mathbb{E}[ q_A q_B ] + \mathbb{E}[ p_A p_B ] & 
    0 & 
    \mathbb{E}[ q_B^2 ] + \mathbb{E}[ p_B^2 ]
    \end{array}
    \right) \, ,
\end{align}
where the matrix entries with the post-holder $\star$ can be assumed to be zero with no loss of generality.

The security of CV QKD, when the state is Gaussian and i.i.d., is well established \cite{Gaussian}. 
Knowing the entries of the above CM, we can bound how many secret bits can distilled from the state.
As a matter of fact, it is sufficient to know upper bounds on the diagonal entries $\mathbb{E}[q_A^2] + \mathbb{E}[p_A^2]$, $\mathbb{E}[q_B^2] + \mathbb{E}[p_B^2]$, and a lower bound on the absolute value of the off-diagonal term $\mathbb{E}[ q_A q_B ] - \mathbb{E}[ p_A p_B ]$.
Given the output of heterodyne detection, $q_{Aj}$, $p_{Aj}$, $q_{Bj}$, $p_{Bj}$, for $j=1,\dots, k$, we can estimate the following bounds:
\begin{align}
\mathbb{E}[q_A^2] + \mathbb{E}[p_A^2] \leq \frac{1}{1-t} \frac{1}{k} \sum_{j=1}^k q_{Aj}^2 + p_{Aj}^2 \, , \label{est-1} \\
\mathbb{E}[q_B^2] + \mathbb{E}[p_B^2] \leq \frac{1}{1-t} \frac{1}{k} \sum_{j=1}^k q_{Bj}^2 + p_{Bj}^2  \, , \label{est-2} 
\end{align}
and
\begin{align}\label{est-3}
\left| \mathbb{E}[q_A q_B] - \mathbb{E}[p_A p_B] \right| & \geq
\frac{1}{1-t^2} \frac{1}{k} \left| \sum_{j=1}^k q_{Aj} q_{Bj} - p_{Aj} p_{Bj} \right| 
- \frac{t}{1-t^2} \frac{1}{k} \sum_{j=1}^k \frac{q_{Aj}^2 + q_{Bj}^2 + p_{Aj}^2 + p_{Bj}^2}{2}
\, .
\end{align}
For any $t  >0$, these bounds hold with probability larger than $1- 8 e^{- k t^2/8}$.
These estimates are obtained in the Appendix \ref{App:tails}.
Although these bounds are not necessarily optimal, the important point here is that they are invariant under the symmetry group 
$q_{Aj} + i p_{Aj} \to \sum_h U_{jh} (q_{Ah} + i p_{Ah})$, 
$q_{Bj} + i p_{Bj} \to \sum_h U_{jh}^* (q_{Bh} + i p_{Bh})$.

\end{widetext}

\subsection{MDI in the PM representation}\label{sec:MDI-PM}

The equivalence of the EB and PM representation follows from (see Section \ref{sec:toolbox} for more detail): 
1) if we measure by heterodyne detection one mode of a TMSV with $N$ mean photon per mode, and obtain the value $\beta$, then the other mode is prepared in a coherent state with amplitude $\alpha = \sqrt{\frac{N}{N+1}} \beta^*$; and
2) if we first displace a mode and then measure it by heterodyne, this is equivalent to first measure and then displace the measurement output.
Therefore, the EB protocol described in Section \ref{sec:MDI_0} is equivalent to the following PM protocol:

\begin{enumerate}[label=\alph*)]

    \item Alice prepares coherent states $|\alpha^0\rangle$ by sampling its complex amplitude $\alpha^0 = (q_A^{\mathrm{pre},0} + i p_A^{\mathrm{pre},0})/\sqrt{2}$ from a circularly symmetric Gaussian distribution with zero mean and variance $N_A$.
    Similarly, Bob prepares coherent states with amplitude 
    $\beta^0 = (q_B^{\mathrm{pre},0} + i p_B^{\mathrm{pre},0})/\sqrt{2}$ sampling from a Gaussian distribution with variance $N_B$.
    They retain the values of the amplitudes, and send the coherent states to the relay.
    
    \item The relay publicly announces a complex number $z = (q_Z + i p_Z)/\sqrt{2}$.
    
    \item Alice and Bob apply the following linear transformation to their local amplitude data:  
    \begin{align}
        \alpha^0 & \to \alpha = \alpha^0 + \sqrt{\frac{N_A}{N_A+1}} \, a z^* \, , \, \, \\
        \beta^0 & \to  \beta = \beta^0 + \sqrt{\frac{N_B}{N_B+1}} \, b z \, .
    \end{align}
    We put $\alpha = (q_A^\mathrm{pre} + i p_A^\mathrm{pre})/\sqrt{2}$ and 
    $\beta = (q_B^\mathrm{pre} + i p_B^\mathrm{pre})/\sqrt{2}$, from which it follows that 
    \begin{align}
    q_A^\mathrm{pre} & = q_A^{\mathrm{pre},0} + a q_Z \, , \\
    p_A^\mathrm{pre} & = p_A^{\mathrm{pre},0} - a p_Z \, , \\
    q_B^\mathrm{pre} & = q_B^{\mathrm{pre},0} + b q_Z \, , \\
    p_B^\mathrm{pre} & = p_B^{\mathrm{pre},0} + b p_Z \, .    
    \end{align}
    
\end{enumerate}
These steps are repeated $n$ times.
Then they proceed as follows:
\begin{enumerate}[label=\alph*)]
  \setcounter{enumi}{3}
  
    \item They perform the resource-efficient energy test as described in Section \ref{sec:test}.

    \item They apply a rotation on their raw keys, $\alpha_j \to \sum_{h=1}^n U_{jh}^* \alpha_h$, $\beta_j \to \sum_{h=1}^n U_{jh} \beta_h$, where $U$ is a $n \times n$ random unitary matrix.
    
    They then select $k < n$ elements (for example $\alpha_1, \dots, \alpha_k$ and $\beta_1, \dots, \beta_k$) and share their values on a public channel. These data allow Alice and Bob to estimate the number of secret bits that can be distilled from what is left of the raw keys.
    
    \item Alice and Bob perform error correction and privacy amplification to distill their secret key from the remaing raw data of size $n-k$.

\end{enumerate}

The security of this PM protocol follows from the one of the EB protocol.
The equivalence between the two representation follows from the equivalence between the corresponding steps 
$a \equiv 1 \& 4$, 
$b \equiv 2$,
$c \equiv 3$.

\subsection{A closer look at parameter estimation}\label{ssec:PE}

In the PM representation of the protocol, Alice and Bob need to estimate the CM $V_\mathrm{PM}^\mathrm{MDI}$ in Eq.\ (\ref{CM-MDI}) from the preparation data $q_{Aj}^\mathrm{pre}$, $p_{Aj}^\mathrm{pre}$, $q_{Bj}^\mathrm{pre}$, $p_{Bj}^\mathrm{pre}$, for $j=1,\dots, n$. 
In turn, they obtain their estimate of the CM $V_\mathrm{EB}^\mathrm{het}$ using Eq.\ (\ref{sdr0i0}). 
To make the notation lighter, below we drop the label $\mathrm{pre}$, and simply refer to the preparation data as
$q_{Aj}$, $p_{Aj}$, $q_{Bj}$, $p_{Bj}$.

The covariant parameter estimation routine relies on the computation of following quantities form a sample of the raw data:
\begin{align}
    C_1 & = \frac{1}{k} \sum_{j=1}^k q_{Aj}^2 + q_{Bj}^2 \, , \\
    C_2 & = \frac{1}{k} \sum_{j=1}^k p_{Aj}^2 + p_{Bj}^2 \, , \\
    C_3 & = \frac{1}{k} \left| \sum_{j=1}^k q_{Aj} q_{Bj} - p_{Aj} p_{Bj} \right| \, .
\end{align}
Note that $C_1$ and $C_2$ can be computed locally by Alice and Bob. However, to compute $C_3$ they need to share their local data on a public channel. This implies that the data used for parameter estimation are compromised and cannot be used for secret key extraction.
Here we show that this limitation can be avoided by using the public data $q_{Zj}$, $p_{Zj}$ from the relay.

In the PM representation of the protocol we have (putting $a' := \sqrt{\frac{N_A}{N_A+1}} \, a$, $b' := \sqrt{\frac{N_B}{N_B+1}} \, b$)
\begin{widetext}
\begin{align}
    C_3 & = \frac{1}{k} \left| \sum_{j=1}^k q_{Aj} q_{Bj} - p_{Aj} p_{Bj} \right| \\
    & = \frac{1}{k} \left| \sum_{j=1}^k ( q_{Aj}^0 + a' q_{Zj} ) ( q_{Bj}^0 + b' q_{Zj} ) - ( p_{Aj}^0 - a' p_{Zj} ) ( p_{Bj}^0 + b' p_{Zj} ) \right| \\
    & = \frac{1}{k} \left| \sum_{j=1}^k q_{Aj}^0 q_{Bj}^0 - p_{Aj}^0 p_{Bj}^0 
    + \frac{1}{k} \sum_{j=1}^k a' \left( q_{Zj} q_{Bj}^0 + p_{Zj} p_{Bj}^0 \right)  
    + b' \left( q_{Aj}^0 q_{Zj} - p_{Aj}^0 p_{Zj} \right)
    + a' b' \left( q_{Zj}^2 + p_{Zj}^2 \right) \right| \, .
\end{align}
We can then write $C_3 = \left| C_{3,1} + C_{3,2} \right|$, where
\begin{align}
    C_{3,1} = \frac{1}{k} \sum_{j=1}^k q_{Aj}^0 q_{Bj}^0 - p_{Aj}^0 p_{Bj}^0 \, ,
\end{align}
and
\begin{align}
    C_{3,2} = \frac{1}{k} \sum_{j=1}^k a' \left( q_{Zj} q_{Bj}^0 + p_{Zj} p_{Bj}^0 \right)  
    + b' \left( q_{Aj}^0 q_{Zj} - p_{Aj}^0 p_{Zj} \right)
    + a' b' \left( q_{Zj}^2 + p_{Zj}^2 \right) \, .
\end{align}

Note that $C_{3,2}$ can be computed locally by Alice and Bob using only their measured data and the public data $q_{Zj}$, $p_{Zj}$.
Consider now the first term,
\begin{align}
    C_{3,1} = \frac{1}{k} \sum_{j=1}^k q_{Aj}^0 q_{Bj}^0 - p_{Aj}^0 p_{Bj}^0 \, .
\end{align}
Note that, by construction of the PM protocol, we know that 
$\mathbb{E}[q_{A}^0 q_{B}^0] = \mathbb{E}[p_{A}^0 p_{B}^0] = 0$,
and
$\mathbb{E}[(q_{A}^0)^2] = \mathbb{E}[(q_{A}^0)^2] = N_A$,
$\mathbb{E}[(q_{B}^0)^2] = \mathbb{E}[(q_{B}^0)^2] = N_B$.
We can then apply the tail bounds (see Appendix \ref{App:tails}) 
\begin{align}
\mathrm{Pr} \left\{ \frac{1}{k} \sum_{j=1}^k q_{Aj}^0 q_{Bj}^0 - p_{Aj}^0 p_{Bj}^0 > 
t (N_A + N_B)  \right\} 
& < 4e^{-k t^2/8} \,  , \\
\mathrm{Pr} \left\{ \frac{1}{k} \sum_{j=1}^k q_{Aj}^0 q_{Bj}^0 - p_{Aj}^0 p_{Bj}^0 < 
- t (N_A + N_B) \right\} 
& < 4e^{-k t^2/8} \,  ,
\end{align}
which imply
\begin{align}
& \mathrm{Pr} \left\{ 
\frac{1}{k} \left| \sum_{j=1}^k q_{Aj} q_{Bj} - p_{Aj} p_{Bj} \right|  < 
\left| C_{3,2} \right| - t (N_A + N_B) \right\}
< 4e^{-k t^2/8} \, .
\label{bound000}
\end{align}

Finally, by combining Eq.\ (\ref{bound000}) with Eqs.\ (\ref{est-1})-(\ref{est-3}), we establish the following bounds on the entries of the CM:
\begin{align}
\mathbb{E}[q_A^2] + \mathbb{E}[p_A^2] \leq \frac{1}{1-t} \frac{1}{k} \sum_{j=1}^k q_{Aj}^2 + p_{Aj}^2 \, , \\
\mathbb{E}[q_B^2] + \mathbb{E}[p_B^2] \leq \frac{1}{1-t} \frac{1}{k} \sum_{j=1}^k q_{Bj}^2 + p_{Bj}^2  \, , 
\end{align}
and
\begin{align}
& \left| \mathbb{E}[q_A q_B] - \mathbb{E}[p_A p_B] \right| \geq 
\frac{1}{1-t^2} \left| C_{3,2} \right| - \frac{t}{1-t^2} (N_A + N_B)  
- \frac{t}{1-t^2} \frac{1}{k} \sum_{j=1}^k \frac{q_{Aj}^2 + q_{Bj}^2 + p_{Aj}^2 + p_{Bj}^2}{2}
\, .
\end{align}
These bounds hold, for any $t>0$, with probability larger than $1 - 12 e^{- k t^2/8}$.

In conclusion, we have shown that in the PM representation of the MDI protocol, parameter estimation can be done locally, with no need to compromise the raw keys. 
In particular, this implies that Alice and Bob can chose $k=n$ and use all the raw keys for both parameter estimation and secret key extraction.

\end{widetext}

\subsection{Resource-efficient CV MDI QKD}\label{sec:debunk}

We are finally in the position of presenting a PM protocol for CV MDI QKD where Alice and Bob can use all their raw data for both key extraction and parameter estimation (as well as the energy test).
The state preparation phase is as for (a)-(c) in Section \ref{sec:MDI-PM}.
The key extraction phase is as follows:
\begin{enumerate}[label=\alph*)]
  \setcounter{enumi}{3}

    \item Alice and Bob apply the resource-efficient energy test as described in Section \ref{sec:test}.

    \item They perform parameter estimation as described in Section \ref{ssec:PE}, with $k=n$. Note that no information about the raw keys is revealed. They can therefore use all the raw data for key extraction. 
    Also note that the active symmetrization is not longer needed.
    
    \item Finally, Alice and Bob perform error correction and privacy amplification to distill their secret key from all the block of raw data of size $n$.

\end{enumerate}

\begin{figure}[t!]
\includegraphics[width=0.9\linewidth]{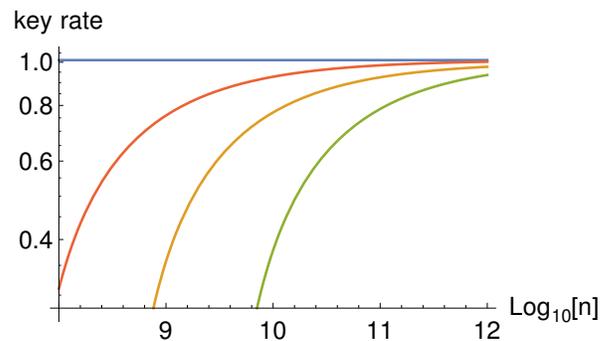}
 \caption{Key rate (against Gaussian attacks) versus the sample size $n$. The top line (blue) shows the asymptotic key rate. The second line from the top (red) is the finite-size key rate obtained with our resource-efficient parameter estimation. The other two lines are for traditional parameter estimation with $k = 10^{-3} n$ (bottom line, green) and $k = 10^{-2} n$ (second line from the bottom, orange).
The plot is obtained assuming a symmetric scenario where the channel from Alice to the relay is the same as from Bob to the relay. 
Such a channel has loss $1 dB$, excess noise of $0.01$ in shot noise units, the input photon number is $10$, the error correction efficiency is $0.95$, and the security parameter is $10^{-20}$.}
\label{fig:rates}
\end{figure}

As an example, Fig.\ \ref{fig:rates} shows a comparison of the finite-size key rates (against Gaussian attacks) obtained using different approaches. 
The plot is obtained for the symmetric setting where the channel from Alice is the same as the channel from Bob.
The choice of the other parameters are described in the figure caption. 
To emphasise the impact of our method, the plot is obtained by taking into account only the finite-size effects related to parameter estimation.
As expected, this numerical example shows that the impact on boosting the key rate can be relevant in the finite-size regime. 
\color{black}

\section{Conclusions}\label{sec:end}

We have analysed the efficiency of two routines in CV QKD: energy test and parameter estimation, the latter applied to a class of MDI protocols.
We have shown that these routines can be realized in a resource-efficient way where all the raw data are used for key extraction, hence improving the expected secret key rate.
{Here we have focused on the no-switching protocols of Weedbrook et al.\ \cite{Weedbrook} and Braunstein \& Pirandola \cite{MDI1}. 
The security analysis of \textit{switching} protocols, where the users choose between two quadratures for encoding information, is less developed when it comes to general, coherent attacks.
However, it is reasonable to expect that our conclusions may as well be extended to these protocols.}

The goal of the energy test is to project the infinite dimensional Hilbert space that characterizes CV systems into a finite dimensional one. Here we have used an approach previously introduced in Ref.\ \cite{Lev2015} in the context of parameter estimation, to show that all the raw data can be used for the energy test and for key extraction.
As the energy test is necessary to achieve composable security against coherent attacks, this result will allow to improve the feasibility of CV QKD.
%
In fact, our approach does not require active symmetrization of the measured data, which is computationally demanding and would make CV QKD impractical.
\color{black}

We have as well analysed the routine of parameter estimation in CV MDI QKD, and have established the composable security of the protocols introduced in Refs.\ \cite{PRL2018,PRA2018}. These protocols have the property that they allow us to use all the raw data for both key extraction and parameter estimation (in some regimes this property can be extended to more general protocols \cite{PRL2018}). 
To show this we have applied the following line of reasoning: 
(1) our starting point is a standard CV MDI QKD protocol, described in the EB representation, whose composable security against coherent attacks has been established in Refs.\ \cite{Lev2017,Ghorai2019}; 
(2) we have considered its equivalent PM protocol, which is also secure; 
(3) we have shown that the parameter estimation in the PM protocol is covariant under the relevant symmetry group and can be implemented without revealing any information about the raw data; 
(4) finally, as no raw data is revealed, it follows that it is possible for Alice and Bob to use all their raw keys for both parameter estimation and key extraction, hence allowing a higher key rate in the finite-size regime.

\begin{acknowledgments}
This work was supported by the EPSRC Quantum Communications Hub, Grant No.\ EP/T001011/1. 
\end{acknowledgments}

\newpage

\appendix

\begin{widetext}

\section{Covariance matrices}\label{App:CM}

Consider a state $\rho_{AB}$ of two modes. The CM of the Wigner function is
\begin{align}
    V_\mathrm{W} = \frac{1}{2}
    \langle \left(\begin{array}{c}
    \hat q_A - \langle \hat q_A \rangle \\
    \hat p_A - \langle \hat p_A \rangle\\
    \hat q_B - \langle \hat q_B \rangle\\
    \hat p_B - \langle \hat p_B \rangle
    \end{array}\right)
\left(
\begin{array}{cccc}
    \hat q_A - \langle \hat q_A \rangle , &
    \hat p_A - \langle \hat p_A \rangle , &
    \hat q_B - \langle \hat q_B \rangle , &
    \hat p_B - \langle \hat p_B \rangle
    \end{array}
    \right) + \mathrm{h.c.} \rangle
\end{align}
where $\mathrm{h.c.}$ stands for the hermitian conjugate, and $\langle O \rangle = \mathrm{Tr}(\hat O \rho_{AB})$ is the quantum mechanical expectation value of the operator $\hat O$.

If we measure the two modes of $\rho_{AB}$ by heterodyne detection, the outcomes of the measurements, $\alpha = (q_A^\mathrm{het} + i p_A^\mathrm{het})/\sqrt{2}$ and $\beta = (q_B^\mathrm{het} + i p_B^\mathrm{het} )\/\sqrt{2}$ define a set of four real-values random variables.
The corresponding CM is
\begin{align}\label{734noi}
    V_\mathrm{EB}^\mathrm{het}= 
    \mathbb{E}\left[ \left(\begin{array}{c}
    q_A^\mathrm{het} - \mathbb{E}[ q_A^\mathrm{het} ] \\
    p_A^\mathrm{het} - \mathbb{E}[ p_A^\mathrm{het} ] \\
    q_B^\mathrm{het} - \mathbb{E}[ q_B^\mathrm{het} ] \\
    p_B^\mathrm{het} - \mathbb{E}[ p_B^\mathrm{het} ]
    \end{array}\right)
\left(\begin{array}{cccc}
    q_A^\mathrm{het} - \mathbb{E}[ q_A^\mathrm{het} ] ,& 
    p_A^\mathrm{het} - \mathbb{E}[ p_A^\mathrm{het} ] , & 
    q_B^\mathrm{het} - \mathbb{E}[ q_B^\mathrm{het} ] , & 
    p_B^\mathrm{het} - \mathbb{E}[ p_B^\mathrm{het} ]
    \end{array}\right) \right] \, ,
\end{align}
where $\mathbb{E}[ O ]$ denotes the expectation value of the random variable $O$.
This is related to the Wigner function CM by the relation
\begin{align}\label{wiem4}
    V_\mathrm{EB}^\mathrm{het} = V_\mathrm{W} + I/2 \, , 
\end{align}
where $I$ is the identity matrix.

Consider now a $1$-way PM protocol, where Alice prepares coherent states of amplitude 
$\alpha = (q_A^\mathrm{pre} + i p_A^\mathrm{pre})/\sqrt{2}$ and sends them to Bob, and Bob measures them by heterodyne detection. 
The amplitude measured by Bob is denoted as $\beta = (q_B^\mathrm{het} + i p_B^\mathrm{het} )/\sqrt{2}$. 
The CM of these variables is 
\begin{align}
    V_\mathrm{PM}^{\text{1-way}} = 
    \mathbb{E}\left[ \left(\begin{array}{c}
    q_A^\mathrm{pre} - \mathbb{E}[ q_A^\mathrm{pre} ] \\
    p_A^\mathrm{pre} - \mathbb{E}[ p_A^\mathrm{pre} ] \\
    q_B^\mathrm{het} - \mathbb{E}[ q_B^\mathrm{het} ] \\
    p_B^\mathrm{het} - \mathbb{E}[ p_B^\mathrm{het} ]
    \end{array}\right)
\left(\begin{array}{cccc}
    q_A^\mathrm{pre} - \mathbb{E}[ q_A^\mathrm{pre} ] ,& 
    p_A^\mathrm{pre} - \mathbb{E}[ p_A^\mathrm{pre} ] , & 
    q_B^\mathrm{het} - \mathbb{E}[ q_B^\mathrm{het} ] , & 
    p_B^\mathrm{het} - \mathbb{E}[ p_B^\mathrm{het} ]
    \end{array}\right) \right] \, ,
\end{align}
which is related to the EB heterodyne CM by the relation
\begin{align}
V_\mathrm{PM}^\text{1-way} = \left(\begin{array}{cc|cc}
\sqrt{\frac{N_A}{N_A+1}} & 0 & 0 & 0 \\
0 & - \sqrt{\frac{N_A}{N_A+1}} & 0 & 0 \\
\hline
0 & 0 & 1 & 0 \\
0 & 0 & 0 & 1
\end{array}\right)
V_\mathrm{EB}^\mathrm{het}
\left(\begin{array}{cc|cc}
\sqrt{\frac{N_A}{N_A+1}} & 0 & 0 & 0 \\
0 & - \sqrt{\frac{N_A}{N_A+1}} & 0 & 0 \\
\hline
0 & 0 & 1 & 0 \\
0 & 0 & 0 & 1
\end{array}\right) \, .
\end{align}

In CV MDI QKD, both Alice and Bob locally prepare coherent states with random amplitudes $\alpha = (q_A^\mathrm{pre} + i p_A^\mathrm{pre})/\sqrt{2}$ and $\beta = (q_B^\mathrm{pre} + i p_B^\mathrm{pre})/\sqrt{2}$, where the amplitudes are Gaussian with zero mean and variance $N_A$ and $N_B$, respectively.
The CM of these variables is
\begin{align}\label{CM-MDI}
    V_\mathrm{PM}^{\mathrm{MDI}} = 
    \mathbb{E}\left[ \left(\begin{array}{c}
    q_A^\mathrm{pre} - \mathbb{E}[ q_A^\mathrm{pre} ] \\
    p_A^\mathrm{pre} - \mathbb{E}[ p_A^\mathrm{pre} ] \\
    q_B^\mathrm{pre} - \mathbb{E}[ q_B^\mathrm{pre} ] \\
    p_B^\mathrm{pre} - \mathbb{E}[ p_B^\mathrm{pre} ]
    \end{array}\right)
\left(\begin{array}{cccc}
    q_A^\mathrm{pre} - \mathbb{E}[ q_A^\mathrm{pre} ] ,& 
    p_A^\mathrm{pre} - \mathbb{E}[ p_A^\mathrm{pre} ] , & 
    q_B^\mathrm{het} - \mathbb{E}[ q_B^\mathrm{pre} ] , & 
    p_B^\mathrm{pre} - \mathbb{E}[ p_B^\mathrm{pre} ]
    \end{array}\right) \right] \, ,
\end{align}
with
\begin{align}\label{sdr0i0}
V_\mathrm{PM}^\mathrm{MDI} = \left(\begin{array}{cc|cc}
\sqrt{\frac{N_A}{N_A+1}} & 0 & 0 & 0 \\
0 & - \sqrt{\frac{N_A}{N_A+1}} & 0 & 0 \\
\hline
0 & 0 & \sqrt{\frac{N_B}{N_B+1}} & 0 \\
0 & 0 & 0 & - \sqrt{\frac{N_B}{N_B+1}}
\end{array}\right)
V_\mathrm{EB}^\mathrm{het}
\left(\begin{array}{cc|cc}
\sqrt{\frac{N_A}{N_A+1}} & 0 & 0 & 0 \\
0 & - \sqrt{\frac{N_A}{N_A+1}} & 0 & 0 \\
\hline
0 & 0 & \sqrt{\frac{N_B}{N_B+1}} & 0 \\
0 & 0 & 0 & - \sqrt{\frac{N_B}{N_B+1}}
\end{array}\right) \, .
\end{align}

\section{Parameter estimation for Gaussian attacks}\label{App:tails}

Consider $k$ i.i.d.\ Gaussian variables $X_1, \dots X_k$, which are with zero mean and variance $\mathbb{E}[X^2]$.
The random variable $\chi_k = \mathbb{E}[X^2]^{-1} \sum_{j=1}^k X_j^2$ is distributed as $\chi^2$ variable of degree $k$.
We have
%
%
\begin{align}
    \mathrm{Pr} \left\{ \frac{1}{k} \sum_{j=1}^k X_j^2 > (1+t) \mathbb{E}[X^2] \right\} < e^{- k t^2/8} \, , \\
    \mathrm{Pr} \left\{ \frac{1}{k} \sum_{j=1}^k X_j^2 < (1-t) \mathbb{E}[X^2] \right\} < e^{- k t^2/8} \, .
\end{align}

Consider now another set of i.i.d.\ zero-mean Gaussian variables $Y_1, \dots Y_k$, with variance $\mathbb{E}[Y^2]$.
From
\begin{align}
    \left\{ \sum_{j=1}^k X_j^2 > a \right\} \mathrm{AND} \left\{ \sum_{j=1}^k Y_j^2 > b \right\} 
    \Rightarrow \left\{ \sum_{j=1}^k X_j^2 + Y_j^2 > a + b  \right\} 
\end{align}
it follows that 
\begin{align}
    \left\{ \sum_{j=1}^k X_j^2 + Y_j^2 < a + b  \right\} 
    \Rightarrow  
    \left\{ \sum_{j=1}^k X_j^2 < a \right\} \mathrm{OR} \left\{ \sum_{j=1}^k Y_j^2 < b \right\} \, .
\end{align}
This in turn implies 
\begin{align}
    \mathrm{Pr} \left\{ \sum_{j=1}^k X_j^2 + Y_j^2 < a + b  \right\} 
    \leq
    \mathrm{Pr}\left\{ \sum_{j=1}^k X_j^2 < a \right\} +
    \mathrm{Pr}\left\{ \sum_{j=1}^k Y_j^2 < b \right\} \, .
\end{align}

In particular, we obtain
\begin{align}
    \mathrm{Pr} \left\{ \mathbb{E}[X^2] + \mathbb{E}[Y^2] > \frac{1}{1-t} \frac{1}{k} \sum_{j=1}^k X_j^2 + Y_j^2 \right\} < 2 e^{- k t^2/8} \, .
\end{align}

\vspace{1cm}

Consider now the identities
\begin{align}
    (X+Y)^2 - (X-Y)^2 & = 4 X Y \, , \\
    (X+Y)^2 + (X-Y)^2 & = 2 X^2 + 2 Y^2 \, ,
\end{align}
where $(X + Y)$ and $(X - Y)$ are both Gaussian variables.
We can therefore write
\begin{align}
\mathrm{Pr} \left\{ \mathbb{E}[(X + Y)^2] < \frac{1}{1+t} \frac{1}{k} \sum_{j=1}^k (X_j + Y_j)^2   \right\} & < e^{-k t^2/8} \, , \\
\mathrm{Pr} \left\{ \mathbb{E}[(X - Y)^2] > \frac{1}{1-t}\frac{1}{k} \sum_{j=1}^k (X_j - Y_j)^2   \right\} & < e^{-k t^2/8} \, .
\end{align}

This implies
\begin{align}
\mathrm{Pr} \left\{ 
\mathbb{E}[(X + Y)^2] - \mathbb{E}[(X - Y)^2]
< \frac{1}{1+t} \frac{1}{k} \sum_{j=1}^k (X_j + Y_j)^2  
- \frac{1}{1-t} \frac{1}{k} \sum_{j=1}^k (X_j - Y_j)^2   \right\}
& < 2 e^{-k t^2/8} \, ,
\end{align}
which is equivalent to
%
\begin{align}
\mathrm{Pr} \left\{ 
\mathbb{E}[XY]
< \frac{1}{1-t^2} \frac{1}{k} \sum_{j=1}^k X_j Y_j  
- \frac{t}{1-t^2} \frac{1}{k} \sum_{j=1}^k \frac{X_j^2 + Y_j^2}{2}
   \right\}
& < 2 e^{-k t^2/8} \, .
\end{align}


In the same way, starting from
\begin{align}
\mathrm{Pr} \left\{ \mathbb{E}[(X + Y)^2] > \frac{1}{1-t} \frac{1}{k} \sum_{j=1}^k (X_j + Y_j)^2   \right\} & < e^{-k t^2/8} \, , \\
\mathrm{Pr} \left\{ \mathbb{E}[(X - Y)^2] < \frac{1}{1+t} \frac{1}{k} \sum_{j=1}^k (X_j - Y_j)^2   \right\} & < e^{-k t^2/8} \, ,
\end{align}
we obtain
%
\begin{align}
\mathrm{Pr} \left\{ 
\mathbb{E}[XY]
> \frac{1}{1-t^2} \frac{1}{k} \sum_{j=1}^k X_j Y_j
+ \frac{t}{1-t^2} \frac{1}{k} \sum_{j=1}^k \frac{X_j^2 + Y_j^2}{2}   
\right\}
& < 2 e^{-k t^2/8}  \, .
\end{align}

Finally, consider a set of four variables, $X$, $Y$, $W$, $Z$. 
By combining the above results we obtain
\begin{align}
\mathrm{Pr} \left\{ 
\left| \mathbb{E}[XY] - \mathbb{E}[WZ] \right|
< \frac{1}{1-t^2} \frac{1}{k} \left| \sum_{j=1}^k X_j Y_j - W_j Z_j \right|
- \frac{t}{1-t^2} \frac{1}{k} \sum_{j=1}^k \frac{X_j^2 + Y_j^2 + W_j^2 + Z_j^2}{2}   
\right\}
& < 4 e^{-k t^2/8}  \, .
\end{align}

\subsection{Further tail bounds}

Proceeding as above, we obtain
\begin{align}
    \mathrm{Pr} \left\{ \sum_{j=1}^k (X_j + Y_j)^2 - (X_j - Y_j)^2 > 
    (1+t) k \mathbb{E}[(X + Y)^2] 
    - (1-t) k \mathbb{E}[(X - Y)^2] \right\} < 2 e^{-k t^2/8}
\, ,
\end{align}
%
which implies
\begin{align}
\mathrm{Pr} \left\{ \frac{1}{k} \sum_{j=1}^k X_j Y_j > \mathbb{E}[X Y] 
+ t \, \frac{ \mathbb{E}[X^2] + \mathbb{E}[Y^2] }{2}  \right\} 
& < 2e^{-k t^2/8} \,  .
\end{align}

Similarly we obtain
\begin{align}
\mathrm{Pr} \left\{ \frac{1}{k} \sum_{j=1}^k X_j Y_j < \mathbb{E}[X Y] 
- t \, \frac{ \mathbb{E}[X^2] + \mathbb{E}[Y^2] }{2}  \right\} 
& < 2e^{-k t^2/8} \,  .
\end{align}

Finally, given four variables, $X$, $Y$, $W$, $Z$, by combining the above results we obtain
\begin{align}
\mathrm{Pr} \left\{ \frac{1}{k} \sum_{j=1}^k X_j Y_j - W_j Z_j > 
\mathbb{E}[X Y] - \mathbb{E}[W Z]
+ t \, \frac{ \mathbb{E}[X^2] + \mathbb{E}[Y^2] + \mathbb{E}[W^2] + \mathbb{E}[Z^2]}{2}  \right\} 
& < 4e^{-k t^2/8} \,  , \\
\mathrm{Pr} \left\{ \frac{1}{k} \sum_{j=1}^k X_j Y_j - W_j Z_j < 
\mathbb{E}[X Y] - \mathbb{E}[W Z]
- t \, \frac{ \mathbb{E}[X^2] + \mathbb{E}[Y^2] + \mathbb{E}[W^2] + \mathbb{E}[Z^2]}{2}  \right\} 
& < 4e^{-k t^2/8} \,  .
\end{align}

\end{widetext}


\begin{thebibliography}{99}

\bibitem{Rev1}
S. Pirandola, U. L. Andersen, L. Banchi, M. Berta, D. Bunandar, R. Colbeck, D. Englund, T. Gehring, C. Lupo, C. Ottaviani, J. Pereira, M. Razavi, J. S. Shaari, M. Tomamichel, V. C. Usenko, G. Vallone, P. Villoresi, P. Wallden,
Advances in Quantum Cryptography,
Adv. Opt. Photon. {\bf 12}, 1012 (2020).

\bibitem{MMosca}
M. Mosca,
Cybersecurity in an Era with Quantum Computers: Will We Be Ready?,
IEEE Security \& Privacy {\bf 16}, 5 (2018).

\bibitem{Lo}
E. Diamanti, H.-K. Lo, B. Qi and Z. Yuan, 
Practical challenges in quantum key distribution,
npj Quantum Information {\bf 2}, 16025 (2016).

\bibitem{Exp1}
G. Zhang, J. Y. Haw, H. Cai, F. Xu, S. M. Assad, J. F. Fitzsimons, X. Zhou, Y. Zhang, S. Yu, J. Wu, W. Ser, L. C. Kwek, and A. Q. Liu,
An integrated silicon photonic chip platform for continuous-variable quantum key distribution,
Nature Photonics {\bf 13}, 839 (2019).

\bibitem{Exp2}
Y. Zhang, Z. Li, Z. Chen, C. Weedbrook, Y. Zhao, X. Wang, Y. Huang, C. Xu, X. Zhang, Z. Wang, M. Li, X. Zhang, Z. Zheng, B. Chu, X. Gao, N. Meng, W. Cai, Z. Wang, G. Wang, S. Yu, and H. Guo,
Continuous-variable QKD over $50$ km commercial fiber,
Quantum Sci. Technol. {\bf 4} 035006 (2019).

\bibitem{Exp3}
Y. Zhang, Z. Chen, S. Pirandola, X. Wang, C. Zhou, B. Chu, Y. Zhao, B. Xu, S. Yu, and H. Guo,
Long-Distance Continuous-Variable Quantum Key Distribution over $202.81$ km of Fiber,
Phys. Rev. Lett. {\bf 125}, 010502 (2020).

\bibitem{Gaussian}
C. Weedbrook, S. Pirandola, R. Garc\'ia-Patr\'on, N. J. Cerf, T. C. Ralph, J. H. Shapiro, S. Lloyd,
Gaussian Quantum Information,
Rev. Mod. Phys. {\bf 84}, 621 (2012).

\bibitem{Garcia}
R. Garc\'ia-Patr\'on and N. J. Cerf, 
Unconditional optimality of Gaussian attacks against continuous variable quantum key distribution,
Phys. Rev. Lett. {\bf 97}, 190503 (2006).

\bibitem{Navascues} 
M. Navascu\'es, F. Grosshans, and A. Ac\'in, 
Optimality of Gaussian attacks in continuousvariable quantum cryptography,
Phys. Rev. Lett. {\bf 97}, 190502 (2006).

\bibitem{Lev2015}
A. Leverrier,
Composable Security Proof for Continuous-Variable Quantum Key Distribution with Coherent States,
Phys. Rev. Lett. {\bf 114}, 070501 (2015).

\bibitem{compo}
M. Ben-Or, Michal Horodecki, D. W. Leung, D. Mayers, J. Oppenheim,
The Universal Composable Security of Quantum Key Distribution,
Theory of Cryptography: Second Theory of Cryptography Conference, TCC 2005, J.Kilian (ed.) Springer Verlag 2005, vol. 3378 of Lecture Notes in Computer Science, pp. 386-406.

\bibitem{Furrer}
F. Furrer, T. Franz, M. Berta, A. Leverrier, V. B. Scholz, M. Tomamichel, R. F. Werner,
Continuous Variable Quantum Key Distribution: Finite-Key Analysis of Composable Security against Coherent Attacks
Phys. Rev. Lett. {\bf 109}, 100502 (2012); 
Phys. Rev. Lett. {\bf 112}, 019902(E) (2014).

\bibitem{FurrerRR}
F. Furrer, 
Reverse-reconciliation continuous-variable quantum key distribution based on the uncertainty principle,
Phys. Rev. A {\bf 90}, 042325 (2014). 

\bibitem{posts}
M. Christandl, R. K\"onig, and R Renner,
Postselection Technique for Quantum Channels with Applications to Quantum Cryptography,
Phys. Rev. Lett. {\bf 102}, 020504 (2009).

\bibitem{Lev2017}
A. Leverrier,
Security of Continuous-Variable Quantum Key Distribution via a Gaussian de Finetti Reduction,
Phys. Rev. Lett. {\bf 118}, 200501 (2017).

\bibitem{Weedbrook}
C. Weedbrook, A. M. Lance, W. P. Bowen, T. Symul, T. C. Ralph, and P. K. Lam,
Quantum Cryptography Without Switching
Phys. Rev. Lett. {\bf 93}, 170504 (2004).

\bibitem{Ghorai2019}
S. Ghorai, E. Diamanti, and A. Leverrier,
Composable security of two-way continuous-variable quantum key distribution without active symmetrization,
Phys. Rev. A {\bf 99}, 012311 (2019).

\bibitem{MDI1}
S. L. Braunstein, S. Pirandola, 
Side-Channel-Free Quantum Key Distribution,
Phys. Rev. Lett. {\bf 108}, 130502 (2012).

\bibitem{MDI2}
H.-K. Lo, M. Curty, B. Qi, 
Measurement-Device-Independent Quantum Key Distribution,
Phys. Rev. Lett. {\bf 108}, 130503 (2012).

\bibitem{MDI-exp1}
Z. Li, Y.-C. Zhang, F. Xu, X. Peng, and H. Guo,
Continuous-variable measurement-device-independent quantum key distribution,
Phys. Rev. A {\bf 89}, 052301 (2014).

\bibitem{MDI-exp2}
S. Pirandola, C. Ottaviani, G. Spedalieri, C. Weedbrook, S. L. Braunstein, S. Lloyd, T. Gehring, C. S. Jacobsen, and U. L. Andersen,
High-rate measurement-device-independent quantum cryptography,
Nature Photonics {\bf 9}, 397 (2015).

\bibitem{PRL2018}
C. Lupo, C. Ottaviani, P. Papanastasiou, and S. Pirandola,
Parameter Estimation with Almost No Public Communication for Continuous-Variable Quantum Key Distribution,
Phys. Rev. Lett. {\bf 120}, 220505 (2018).

\bibitem{PRA2018}
C. Lupo, C. Ottaviani, P. Papanastasiou, and S. Pirandola,
Continuous-variable measurement-device-independent quantum key distribution: Composable security against coherent attacks,
Phys. Rev. A {\bf 97}, 052327 (2018).

\bibitem{Pirs2008}
S. Pirandola, S. Mancini, S. Lloyd, and S. L. Braunstein,
Continuous-variable quantum cryptography using two-way quantum communication,
Nat. Phys. {\bf 4}, 726 (2008).

\bibitem{QZ2018-2way}
Q. Zhuang, Z. Zhang, N. L\"utkenhaus, and J. H. Shapiro,
Security-proof framework for two-way Gaussian quantum-key-distribution protocols,
Phys. Rev. A {\bf 98}, 032332 (2018).

\bibitem{QZ2016}
Q. Zhuang, Z. Zhang, J. Dove, F. N. C. Wong, and J. H. Shapiro,
Floodlight quantum key distribution: A practical route to gigabit-per-second secret-key rates,
Phys. Rev. A {\bf 94}, 012322 (2016).

\bibitem{QZ2017}
Z. Zhang, Q. Zhuang, F. N. C. Wong, and J. H. Shapiro,
Floodlight quantum key distribution: Demonstrating a framework for high-rate secure communication,
Phys. Rev. A {\bf 95}, 012332 (2017).

\bibitem{QZ2018}
Q. Zhuang, Z. Zhang, and J. H. Shapiro,
High-order encoding schemes for floodlight quantum key distribution,
Phys. Rev. A {\bf 98}, 012323 (2018).

\bibitem{Ferraro}
A. Ferraro, S. Olivares, M. G. A. Paris,
Gaussian States in Quantum Information
(Bibliopolis, Napoli, 2005).

\bibitem{Aniello},
P. Aniello, C. Lupo, and M. Napolitano,
Exploring Representation Theory of Unitary Groups via Linear Optical Passive Devices,
Open Syst. Inf. Dyn. {\bf 13}, 415 (2006).

\bibitem{GdeFinetti}
A. Leverrier,
$\mathrm{SU(p,q)}$ coherent states and a Gaussian de Finetti theorem,
J. Math. Phys. {\bf 59}, 042202 (2018).



\end{thebibliography}
\end{document}